
\input phyzzx

\Pubnum{ \vbox{\hbox{UM-P-93/75} \hbox{hep-th/9402079}} }
\pubtype{}
\date{February 1994}

\titlepage

\title{The Solutions of Affine and Conformal Affine Toda Field Theories}

\author{G. Papadopoulos\foot{E-mail: gpapas@vxdesy.desy.de}}
\address{II. Institute for Theoretical Physics \break Luruper Chaussee
149\break
         22761 Hamburg Germany}
\andauthor{B. Spence\foot{E-mail: spence@tauon.ph.unimelb.edu.au}}
\address{School of Physics\break University of Melbourne\break Parkville
          3052 Australia}

\abstract{We give new formulations of the solutions of
the field equations of the affine Toda and conformal affine Toda
theories on a cylinder and two-dimensional Minkowski space-time.
These solutions are parameterised in terms of initial data and the resulting
covariant phase spaces are diffeomorphic to the Hamiltonian ones.
We derive the fundamental Poisson brackets of
the parameters of the solutions and
give the general static solutions for the affine theory.}
\endpage

\endpage



\def\RN{{\cal R}}

\def\G {\bf {g}}

\def\hi {\chi}
\def\half{{1\over2}}

\def\pbr#1#2{ \{#1,#2\} }
\def\pl{Phys. Lett.\ }

\def\np {Nucl. Phys.\ }

\def\exp{{\rm exp}}

\def\dpl{\partial_+}
\def\dmi{\partial_-}

\def\intx{\int_0^1\!dx\,}
\def\l {\lambda}

\def\Gp{{\cal L}^+(\G)}
\def\Gm{{\cal L}^-(\G)}

\def\a{\alpha}

\def\H{{\cal H}}

\def\k{\kappa}
\def\Cab{C_{ij}}
\def\Kab{K_{ij}}

\def\pa{\phi^i}
\def\pb{\phi^j}

\def\Sumd{\sum_{i}}

\def\Lra{\big\vert\Lambda_i\big>}
\def\Lla{\big<\Lambda_i\big\vert}

\def\Lrrr{\big\vert\Lambda_r\big>}
\def\Llrr{\big<\Lambda_r\big\vert}
\def\Lro{\big\vert\Lambda_{{}_0}\big>}
\def\Llo{\big<\Lambda_{{}_0}\big\vert}

\def\W {{\cal {W}}}


\REF\ls {A.N. Leznov and M.V. Saveliev, Lett. Math. Phys. {\bf 3} (1979)
      489; Commun. Math. Phys. {\bf 74} (1980) 111; Lett. Math. Phys. {\bf 6}
(1982) 505;
      Commun. Math. Phys. {\bf 83} (1983) 59; J. Sov. Math. {\bf 36} (1987)
699;
    Acta Appl. Math. (1989) 1.}
\REF\oliveturok{D.I. Olive and N. Turok, \np {\bf B220} (1983) 491, {\bf B257}
(1985)
       277, {\bf B265} (1986) 469.}
\REF\mansfield{P. Mansfield, \np {\bf B208} (1982) 277, {\bf B222} (1983) 419.}
\REF\us{G. Papadopoulos and B. Spence, \pl {\bf B295} (1992) 44; \pl {\bf B308}
(1993) 253.}
\REF\usthree{G. Papadopoulos and B. Spence, Class. Quantum Gravity, in press.}
\REF\intpert{A.B. Zamolodchikov, Int. J. Mod. Phys. Lett. {\bf A1} (1989)
4235;\hfil\break
             T. Eguchi and S.-K. Yang, \pl {\bf B224} (1989) 373;\hfil\break
            T.J. Hollowood and P. Mansfield, \pl {\bf B226} (1989) 73.}
\REF\atsmatrix{H.W. Braden, E. Corrigan, P.E. Dorey and R. Sasaki, \np {\bf
B338} (1990)
        689, {\bf B356} (1991) 469; P.E. Dorey, \np {\bf B358} (1991) 654,
         {\bf B374} 1992) 74;\hfil\break
         T.R. Klassen and E. Melzer, \np {\bf B338} (1990) 485;\hfil\break
          M.D. Freeman, \pl {\bf B261} (1991) 57;\hfil\break
          A. Fring, H.C. Liao and D.I. Olive, \pl {\bf B266} (1991) 82;
           A. Fring, G. Mussardo and P. Simonetti, Imperial preprint
TP/91-92/31.}
\REF\soliton{T.J. Hollowood, \np {\bf B384} (1992) 523;
         \hfil\break D.I. Olive, N. Turok and J. Underwood,
        {\it Affine Toda Solitons and Vertex Operators},
hep-th/9305160;\hfil\break
        J. Underwood, {\it Aspects of Non-Abelian Toda Theory},
hep-th/9304156;\hfil\break
       M. Kneipp and D.I. Olive, {\it Crossing and Anti-Solitons},
hep-th/9305154.}
\REF\otu{D.I. Olive, N. Turok and J. Underwood, \np {\bf B362} (1993) 294.}
\REF\catpapers{O. Babelon and L. Bonora, \pl {\bf B244} (1990) 220;
        \hfil\break L. Bonora, M. Martellini and Y.-Z. Zhang, \pl {\bf B253}
(1991) 373.}

\sequentialequations


\chapter{Introduction}
        It has been known for many years that the field equations of certain
two-dimensional field theories can be solved exactly.  Some of these theories
are the Wess-Zumino-Witten model, Liouville field theory and and the various
versions of Toda field theory.  The field equations of these theories are
non-linear, partial, hyperbolic differential equations and
various discussions of their solutions have been presented in the literature.
One method for
solving such integrable systems is based on the theory of Lax pairs and this
has been extensively studied for Toda systems [\ls -\mansfield].  These authors
solve the field equations on two-dimensional Minkowski space-time
and describe their solutions in terms of functions that depend on the
light-cone co-ordinates $x^{\pm}$ of the Minkowski space-time.  Some
difficulties with these solutions are that
the ranges of the solution parameters are not clearly specified, and the
relation
of the parameters to the initial data of these theories is not given.

To deal with the above deficiencies,  new parameterisations of solutions of the
field equations of the Wess-Zumino-Witten, Liouville, Toda and non-Abelian Toda
conformal field theories were given in refs. [\us -\usthree].
These parameterisations have the advantage that the
solutions of these theories are explicitly
well-defined for all the values of their
parameters, the Poisson brackets of their parameters can be directly
calculated, and the
spaces of these parameters are isomorphic to the spaces of the associated
initial data, \ie\ the solutions of these theories are parameterised directly
in terms of their initial data. This made it possible to give covariant phase
space descriptions of these theories which are explicitly diffeomorphic to the
Hamiltonian phase space formulations.

A lot of attention has been focused recently on integrable, but not necessarily
conformal, two-dimensional field theories.  An important class  is the affine
Toda field theories, which are integrable perturbations of conformal field
theories
[\intpert].  These models have a
rich algebraic structure, involving exact S-matrices
and soliton solutions (see refs. [\atsmatrix -\otu] and references therein).
The solutions of the classical field equations of these
theories were discussed recently in ref. [\otu], following the Leznov-Saveliev
description.

In this paper, we will give a new formulation of the solutions of
affine Toda field theory.  This formulation will be along the lines of
similar solution space
parameterisations given by us for the WZW and Toda-type field theories.
In this new parameterisation, the affine Toda
solutions are well-defined over all space, which is taken to be a cylinder
or two-dimensional Minkowski space-time.  The Poisson
brackets of the parameters of the solutions are calculated. An explicit
isomorphism is also constructed between the space of initial data and the space
of parameters of the solutions of affine Toda field theory.  A similar
formulation of the solutions of the
field equations of the conformal affine Toda field theory is also given.
Finally, using our new
approach, the general static solution of the affine Toda theory is presented.


\chapter{The Solutions of Affine Toda Field Theory}

Let $\G$ be  a (semi-) simple  Lie algebra and $\H$ be a Cartan
subalgebra of $\G$.  We introduce a Chevalley basis $(H_i, E_{\a^i},
E_{-\a^i})$ in  $\G$, where $\Delta\equiv \{\a^i, i=1, \dots,l={\rm rank}
\ \G \}$ is the set of simple roots, $H_i\equiv {2\a^i\cdot H \over
\vert\a^i\vert^2}$, $H\in \H$, $E_{ \pm\a^i}$ are the step operators for the
simple roots and $[H_i, H_j]=0$, $[E_{\a^i}, E_{-\a^i}]= H_i$ and $[H_i,
E_{\pm \a^j}]=\pm K_{ji}  E_{\pm \a^j}$ (with no summation over $j$).  The
matrix $K\equiv \{K_{ij}\}$ is the Cartan matrix of $\G$, i.e. $K_{ij}= {2
\a^i\cdot\a^j \over \vert\a^j\vert^2}$.  The symbols $\Phi^+$ ($\Phi^-$)
will denote the sets of positive (negative) roots, respectively, and
$\Phi\equiv \Phi^+\cup \Phi^-$ is the space of all roots of $\G$. We will
also use the symbols $\Gp$ and $\Gm$ to denote the sets of step operators
for the positive and negative roots respectively.  Finally, we normalise
the Killing form $(\cdot, \cdot)$ as follows: $(H_i, H_j)\equiv \Tr(H_i\cdot
H_j) =C_{ij}$,
$(E_{\a^i}, E_{-\a^j})\equiv \Tr(E_{\a^i}\cdot E_{-\a^j}) =
{2\over\vert\a^i\vert^2}\delta_{ij}$ and
$(E_{\a^i}, H_j)\equiv\Tr(E_{\a^i}\cdot H_j) = 0$, where $C_{ij}\equiv
{2\over\vert\a^i\vert^2}
K_{ij}$.

The Lagrangian of affine Toda field theory is
$$
L=-{\k^2\over 8\pi} \Big((\partial_+\phi, \dmi\phi ) + {2 M^2\over \k^2}
    \Big(\Sumd {m^i\over \vert\a^i\vert^2} \exp\big[{\k\over2}
        (\alpha^i)^2 (H_i,\phi)\big] + {1\over \psi^2}
           \exp\big[-{\k\over 2} \psi^2 (m, \phi)\big] \Big)\Big),
\eqn\aaone
$$
or equivalently in components
$$
L=-{\k^2\over 8\pi} \Big(C_{ij}\partial_+\phi^i\dmi\phi^j + {2 M^2\over \k^2}
\Big(\Sumd {m^i\over \vert\a^i\vert^2} \exp\big[\k \Kab\pb\big] + {1\over
\psi^2}
\exp\big[-\k \psi^2 C_{ij} {m^i\over 2} \phi^j\big] \Big)\Big),
\eqn\aone
$$
where $\phi$ is a map from a
cylinder $S^1\times \RN$ to  $\RN^l$ ($\phi \equiv \phi^i H_i$) and $\k$, $M$
are non-zero coupling constants. The symbol $\psi$ denotes the highest root of
$\G$, the integers $m^i$ being defined by the relation
${\psi\over \psi^2}=\Sumd m^i {\a^i\over \vert\a^i\vert^2}$, with $m\equiv m^i
H_i$.  The pairs $(x,t): 0\leq x<1,
-\infty<t<\infty$ are the co-ordinates of $S^1\times \RN$ and
$x^\pm = x \pm t, \partial_\pm = \half(\partial_x \pm\partial_t)$.

The equations of motion following from the Lagrangian \aone\ are
$$
     \dpl\dmi\pa - {M^2 m^i\over 2\k} \Big( \exp\big(\k \Kab\pb\big)-
          \exp \big(-{\k\over2} \psi^2 C_{kl} m^k \phi^l\big)\Big)=0.
       \eqn\atwo
$$

Let $\hat{\G}$ be the affine Lie algebra associated to the Lie algebra $\G$.
We denote by $a^r, r=0, \dots, l$, and $\Lambda_r, r=0, \dots, l$ the simple
roots and the lowest fundamental weights of $\hat{\G}$ respectively.  The
step operators for the simple roots of $\hat {\G}$ are denoted by $\hat
E_{a^r}$.
The new formulation of the solutions of the affine Toda
field equations \atwo\ takes the form
$$
\exp\big(-\kappa \phi^i(x,t)\big) = \exp\big(-\kappa \phi^i_{{}_R}(x^-)\big)
    \;{ \Lla \W (A;x^+,x^-)\Lra
      \over \Llo\W (A;x^+,x^-)\Lro^{m_i}},
\eqn\athree
$$
where $\W$ is the holonomy of a connection $A$ and $\phi_{{}_R}$ is a periodic
function on the real line.
The components of the connection $A\equiv A_{{}_0}+A_{<0} +A_{>0}$ are
$$
\eqalign{
 A_{{}_0}&= \k u,
\cr
A_{<0}&= \mu {\hat E}_{-1},
\cr
A_{>0}&= \nu \exp ({\k \phi_{{}_R}}) {\hat E}_1 \exp({-\k \phi_{{}_R}}), \cr}
\eqn\afour
$$
where $u$ is a periodic one-form on the real line with values in the Cartan
subalgebra $\H$ of $\G$, $\hat E_{\pm 1}\equiv {\sum}^l_{r=0} {\sqrt {m^r}}
\hat E_{\pm a^r}$ and $\mu, \nu$ are non-zero real constants satisfying
the relation $M^2 = 2\mu\nu$.
That the expression for $\phi$ in equation \athree\ solves the
field equations \atwo\ of affine Toda theory follows from exactly the same
argument as that presented in ref. [\otu], where the Leznov-Saveliev
formulation
of the solutions was considered. Our notation follows this reference. We will
not repeat
this argument here.
The periodicity of the solution $\phi$ in \athree\ follows from the periodicity
of its independent parameters $u$ and $\phi_{{}_R}$.  The space of independent
parameters of the solutions is diffeomorphic to $T^*L\RN^l$, i.e the cotangent
bundle of the loop space of $\RN^l$.
Note that the solutions of the affine Toda field theory
on two-dimensional Minkowski
space-time are also given by equation \athree.  The only difference in this
case is that the independent parameters $u$ and $\phi_{{}_R}$ of the solutions
are functions on the real line which are not necessarily periodic.

The Lagrangian symplectic form of the affine Toda theory is
$$
    \Omega = {\k^2\over 16\pi} \intx\Cab\,\delta\pa\,\partial_t\delta\pb,
\eqn\afive
$$
evaluated at $t=0$, where the map $\phi$ in eqn. \afive\ satisfies
the affine Toda field equations of motion \athree. To express $\Omega$ in terms
of the
independent parameters of the solutions, we insert
the Toda solution \athree\ into the form $\Omega$, giving
$$
   \Omega = -{\k^2 \over 16\pi}\intx
\Cab\big(\delta\pa_{{}_R}\,\partial_x\delta\phi^j_{{}_R} -2\delta\pa_{{}_R}\,
        \delta u^j \big).      \eqn\asix
$$
where $u^i=-<\l_i|u|\l_i>$ and $\{\lambda^i; i=1,\dots,l\}$  is the set of
fundamental lowest weights of $\G$.  The Poisson brackets are easily obtained
from eqn. \asix, and are
$$
  \eqalign{ \pbr{\pa_{{}_R}(x)}{\pb_{{}_R}(y)} & = 0, \crr
          \pbr{\pa_{{}_R}(x)}{u^j(y)} & ={8 \pi\over \k^2}
(C^{-1})^{ij}\delta(x,y),\crr
          \pbr{u^i(x)}{u^j(y)} & = {8 \pi\over \k^2}(C^{-1})^{ij}
                                     {\partial_x}\delta(x,y).\crr}
     \eqn\aseven
$$
Note that these brackets are the Poisson brackets
on the cotangent bundle of the loop space of   $\RN^l$.

 The space of initial data of affine Toda field theory is the space of affine
Toda fields $\phi$, at $t=0$, and their time derivatives
${\partial\pa\over\partial t}$ at
$t=0$, and it can be equipped with the symplectic form induced from eqn.
\afive.
It is straightforward to see that the Hamiltonian phase space of affine Toda
field theory is isomorphic as symplectic manifold to the space of initial
data of this theory. There is also an isomorphism between the space of initial
data of affine Toda field theory and the space of parameters of
the solutions \athree\ of the field equations \atwo. From the
solutions \athree\ of affine Toda theory, we find immediately
that the map between the initial data $f^i(x) \equiv \pa(x,0)$, $w^i(x)
\equiv {\partial\pa\over\partial t}(x,0)$ and our covariant phase space
parameters $\phi^i_{{}_R}, u^i$ is
$$
   \eqalign { f^i(x) &= \pa_{{}_R}(x), \cr
            w^i(x)  &= - \partial_x\pa_{{}_R}(x)+2 u^i(x). \cr}  \eqn\aeight
$$
This map is clearly a diffeomorphism. Moreover it is a symplectic
diffeomorphism from the space of initial data to the space of parameters of the
solutions of the affine Toda theory because,   as
it is easy to show, it maps the symplectic form \athree\ to the
form $\Omega = {\k^2 \over 16\pi}\intx \Cab\delta f^i(x)\,\delta w^j(x)$.
Finally, we can use the same argument as in the case of Toda theory of ref.
[\usthree] to prove that the solutions \athree\ of the affine Toda field
theory are real and well-defined for all the values of the parameters
$\phi_{{}_R}$ and $u$.


\chapter { The Conformal Affine Toda Field Theory}

Let $d\equiv -L_{{}_0}$ be the derivation associated with the affine Lie
Algebra $\hat {\G}$.  We denote by $k$ the central charge generator and
extend the inner product $(\cdot,\cdot)$ from $\G$ to the vector space
${\G}_{{}_0}\equiv {\G}\oplus \RN [k,d]$ by taking $(d,k)=1$, $(d,d)=0$,
$(k,k)=0$, $\G$ orthogonal to $\RN[k,d]$, and $\H_{{}_0}\equiv\H\oplus
\RN[k,d]$.  If  $V_1, V_2 \in { {\H}}_{{}_0}$ and $V_1=v_1+v'_1 k +\tilde {v}_1
d$, $v_1\in \H$, similarly for $V_2$, we define $V_1(V_2)\equiv (v_1, v_2) +
v'_1 v'_2 + \tilde {v}_1 \tilde {v}_2$.

The Lagrangian of conformal affine Toda field theory is [\catpapers]
$$
L=-{\k^2\over 8\pi} \Big(\big(\partial_+\Phi, \dmi\Phi \big) + {2 M^2\over
\k^2}
\sum^{l}_{r=0} {m^r\over \vert a^r\vert^2} \exp\big(\k a^r (\Phi)\big) \Big),
\eqn\bone
$$
where $\Phi$ is a map from a cylinder $S^1\times \RN$ to  $\H_0$,
${\hat{\Delta}} \equiv\{a^r; r=0,\dots,l\}$ is the set  of simple roots of the
affine Lie algebra $\hat {\G}$, the positive integers $\{m^r;  r>0\}$ are
defined as in the previous section, $m^{{}^0}=1$, and $\k$ and $M$ are real,
non-zero coupling constants. The pairs $(x,t): 0\leq x<1, -\infty<t<\infty$
are the co-ordinates of $S^1\times \RN$ and
$x^\pm = x \pm t, \partial_\pm = \half(\partial_x \pm\partial_t)$.

The field equations of conformal affine Toda theory following from the
Lagrangian \bone\ are
$$
     \dpl\dmi\Phi - {M^2\over \k} \sum^{l}_{r=0} {m^r H^{a_r}\over \vert
a^r\vert^2} \exp(\k a^r (\Phi))=0.
       \eqn\btwo
$$

It is known that the above field equation can be truncated to the field
equation of affine Toda field theory [\catpapers].
To do this, we rewrite the field
$\Phi$ in the basis $\{H, k, d'=h d+ \theta H\}$ as
$$\Phi=\phi H+ \xi k+ \eta d',
\eqn\bbtwo$$
where $h$ is the Coxeter number of $\G$ and $\theta=\sum^{l}_{i=1}
{\lambda^i\over \vert\a^i\vert^2}$, we
identify the affine Toda field with the first component $\phi$ of $\Phi$, and
set the
fields $\eta$ and $\xi$ to zero.

The solutions of the field equations \btwo\ of conformal affine Toda field
theory
are
$$ \eqalign{
 \exp\big(\kappa \Lambda^r (\Phi(x,t))\big) &= \exp\big(\kappa \Lambda^r
(\Phi_R(x^-))\big)
   \;\Llrr \W (A;x^+,x^-)\Lrrr
    \cr
      \eta&=\int^{x^+}_{x^-}\!\! b(s)\, ds + \eta_R(x^-),\cr}
\eqn\bthree
$$
where $\W$ is the holonomy of a connection $A$, $\Phi_R$ is a map from the real
line into $\H_{{}_0}$, $\{\eta_{{}_R}, b\}$ are maps from the real line into
the real line $\RN$  and $\{\Lambda_r; r=0,\dots, \rm {rank} \G\}$ are the
lowest fundamental weights of the affine Lie algebra $\hat{\G}$.  Note that
$\eta$ in eqn. \bthree\ is a free field, i.e. satisfies
$\partial_+\partial_-\eta=0$.  The addition of the equation for $\eta$ in
\bthree\ is necessary because the lowest fundamental weights $\Lambda_r$ span a
subspace of  $\H_{{}_0}$ of co-dimension one. The components of the connection
$A\equiv A_{{}_0}+A_{>0} +A_{<0} $ in equation \bthree\ are
$$\eqalign{
 A_{{}_0}&= \k u,
\cr
A_{>0}&= \mu {\hat E}_1,
\cr
A_{<0}&= \nu e^{\k \Phi_R} {\hat E}_{-1} e^{-\k\Phi_R},  \cr}
\eqn\bfour
$$
where $u$ is a periodic map from the real line into $\H_{{}_0}$.  The
independent parameters of the solutions of the conformal affine Toda field
theory are $\{u^r\equiv -\Llrr u \Lrrr; r=0,\dots,l\}$, $\{\Phi^r_R\equiv -
\Lambda^r(\Phi_R); r=0,\dots,l\}$, $b$, and $\eta_{{}_R}$, and the space of
independent parameters is diffeomorphic to $T^*L\H_{{}0}$. Note the
difference in the definitions of $u^r$ and $\Phi^r_R$.  The periodicity of the
solutions $\Phi$ of the conformal affine Toda theory in the co-ordinate $x$
follows from the periodicity of the independent parameters $u^r, b, \Phi^r_R$
and $\eta_{{}_R}$. If we consider the conformal affine Toda field theory on
two-dimensional Minkowski space-time, rather than on the cylinder, then
the solutions of this theory are still given
by equation \bthree, but in this case the independent parameters of the
solutions are functions on the real line which are not necessarily periodic.

The Lagrangian symplectic form of the conformal affine Toda theory is
$$
    \Omega = {\k^2\over 16\pi} \intx\,(\delta\Phi\,,\partial_t\delta\Phi),
\eqn\bfive
$$
evaluated at $t=0$, where the map $\Phi$ in eqn. \afive\ satisfies
the field equations \btwo\ of conformal affine Toda theory.
To express the symplectic
form $\Omega$ in terms of the free parameters $u^r$, $\Phi^r_R$, $b$ and
$\eta_{{}_R}$ of the solutions \bthree, we insert these solutions
into $\Omega$, giving
$$
    \Omega = -{\k^2\over 16\pi} \intx\,\big( C_{ij}\, (\delta\varphi_{{}_R}^i\,
\partial_x\delta\varphi_{{}_R}^j - 2  \delta\varphi_{{}_R}^i\, \delta{\tilde
u}^j) +D_{pq}\, (\delta \pi_{{}_R}^p\, \partial_x \delta \pi_R^q - 2  \delta
\pi_{{}_R}^p\, \delta{\tilde b}^q)\big),
 \eqn\bsix
$$
where ${\tilde u}^i= u^i-m^i u^{{}^0}$, $\pi_{{}_R}=(\xi_{{}_R}, \eta_{{}_R})$,
${\tilde {b}}=({2\over \psi^2}u^{{}^0}, b)$, the non-zero components of the
matrix $D$ are $D_{01}=D_{10}=h$, and $p,q=0,1$.  Here we have used the
decomposition $\Phi_R=\varphi_{{}_R}^i H_i+\xi_{{}_R}k+ \Phi^0_R d$ of the
vector $\Phi_R\in \H_{{}_0}$ and $\{u^r;r=0,\dots,l\}=\{u^{{}^0},
u^i;i=1,\dots,l\}$.  Note that $\Phi^0_R$ does not appear in the symplectic
form \bsix.

The symplectic form \bsix\ can be inverted to calculate the Poisson brackets of
the theory. We find them to be
$$\eqalign{ \pbr{\varphi_{{}_R}(x)}{\varphi_{{}_R}(y)} & = 0, \crr
          \pbr{\varphi^i_{{}_R}(x)}{{\tilde u}^j(y)} & ={8 \pi\over \k^2}
(C^{-1})^{ij}\delta(x,y),\crr
          \pbr{{\tilde u}^i(x)}{{\tilde u}^j(y)} & = {8 \pi\over
\k^2}(C^{-1})^{ij} {\partial_x}\delta(x,y), \cr} \quad
\eqalign{
\pbr{\pi_{{}_R}(x)}{\pi_{{}_R}(y)} & = 0, \crr
\pbr{\pi^p_{{}_R}(x)}{{\tilde b}^q(y)} & ={8 \pi\over \k^2}
(D^{-1})^{pq}\delta(x,y),\crr
\pbr{{\tilde b}^p(x)}{{\tilde b}^q(y)} & = {8 \pi\over \k^2}(D^{-1})^{pq}
{\partial_x}\delta(x,y). \cr}
\eqn\bseven
$$
These Poisson brackets can be written in a form similar to that obtaining
for the affine Toda theory
of the previous section, by defining $\hi_{{}_R}\equiv(\varphi_{{}_R},
\pi_{{}_R})$ and $z\equiv ({\tilde u}, {\tilde b})$.  Expressing the above
Poisson brackets in terms of $\hi_{{}_R}$ and $z$, we get
$$
\eqalign{ \pbr{\hi^I_{{}_R}(x)}{\hi^J_{{}_R}(y)} & = 0, \crr
          \pbr{\hi^I_{{}_R}(x)}{{z}^J(y)} & ={8 \pi\over \k^2} ({\hat
{C}}^{-1})^{IJ}\delta(x,y),\crr
          \pbr{{ z}^I(x)}{{z }^J(y)} & = {8 \pi\over \k^2}({\hat
{C}}^{-1})^{IJ} {\partial_x}\delta(x,y),\cr}
\eqn\beight
$$
where the matrix ${\hat C}\equiv C\oplus D$ is the inner product
$(\cdot,\cdot)$ of $G_{{}_0}$ restricted on $\H_{{}_0}$ and evaluated in the
basis $\{H_i, k, d\}$ of $\H_{{}_0}$, and $I,J=1, \dots, {\rm {dim}}\H_{{}_0}$.
It can be shown as in the Toda case
that the space of independent parameters of the
solutions \bthree\ of the field equations of  the conformal affine Toda field
theory is isomorphic as a symplectic manifold  to the space of initial data of
this theory, and that the solutions of this theory are real and
well-defined for all the values of these parameters.


\chapter {Affine Toda Particles and Static Solutions}

The static solutions of the field equations of affine Toda field theory are
those that are independent of the time co-ordinate $t$ of the space-time.  The
static solutions of affine Toda field theory obey the ordinary differential
equation
$$
   \partial_x^2\pa - {2 M^2 m^i\over  \k} \Big( \exp\big(\k \Kab\pb\big)- \exp
\big(-\k
\psi^2 C_{kl} {m^k\over 2} \phi^l\big)\Big)=0,
\eqn\cone
$$
which follows from setting $\partial_t\phi=0$ in
the field equation of affine Toda field theory \atwo.
One way to find the solutions of this equation is to use the formula \athree\
for all solutions of affine Toda field theory, and find which solutions are
independent of $t$.  However, this turns out to be a rather complicated
algebraic problem.  Instead, it is easier to solve equation \cone\ directly
by comparing it with the equation of the affine Toda {\it particle}, which is
$$
\partial_t^2\pa + {2 M^2 m^i\over  \k} \Big( \exp\big(\k \Kab\pb\big)- \exp\big
(-\k \psi^2
C_{kl} {m^k\over 2} \phi^l\big)\Big)=0.
\eqn\ctwo
$$
The solutions of equation \ctwo\ are those of equation \athree\ with the
parameters $\phi_{{}_R}$ and $u$ constant. These are the most
general solutions of this equation and the space of parameters of the solutions
is isomorphic as a symplectic manifold to the cotangent bundle of $\RN^l$.

Next observe that if we transform $t\rightarrow x$ and $M^2\rightarrow -M^2$ in
equation \ctwo, it becomes equation \cone.  To solve equation \cone, we must
find a way to change the overall sign of the potential term in equation \ctwo.
But we know that $M^2=2\mu\nu$, so this can be achieved by setting either
$\mu\rightarrow -\mu$ or $\nu\rightarrow -\nu$ in the definition of the
components of the connection $A$ in equation \afour. Choosing the
former, we deduce that the solutions of
equation \cone\ are
$$
\exp\big(-\kappa \phi^i(x)\big) = \exp\big(-\kappa \phi^i_{{}_R}\big)
     { \Lla \exp (2x {\hat A})
         \Lra \over \Llo\exp (2x {\hat A})\Lro^{m_i}}
\eqn\cthree
$$
where $\phi_{{}_R}$ is a constant parameter and the components of ${\hat A}$
 (${\hat A}\equiv {\hat A}_{{}_0}+{\hat A}_{<0} +{\hat A}_{>0}$) are
$$
\eqalign{   {\hat A}_{{}_0}&= \k u, \quad u\in \H
   \cr
       {\hat A}_{<0}&= -\mu {\hat E}_{-1},
    \cr
         {\hat A}_{>0}&= \nu\, \exp (\k \phi_{{}_R})
            {\hat E}_1 \exp(-\k \phi_{{}_R}). \cr}
\eqn\cfour
$$
Note the sign difference in the definitions of ${\hat A}_{<0}$  (eqn. \cfour)
and $A_{<0}$  (eqn. \afour). The independent parameters of the solutions
\cthree\ are
$\phi_{{}_R}$ and $u$ and the space of parameters is diffeomorphic to the
cotangent bundle of $\RN^l$.  Although there is a natural symplectic
structure on this space, it is not the one induced by the symplectic structure
on
the space of all solutions of the affine Toda field theory studied in
section two.


\noindent{\bf Acknowledgements}\hfil\break
\noindent G.P. was funded by a grant from the European Union and would like to
thank
H. Nicolai for support.
B.S. was supported by a QEII Fellowship from the Australian Research Council,
and
acknowledges the hospitality of the Theory Group at Queen Mary \& Westfield
College
London, where part of this work was carried out.

\refout

\end